\documentstyle{article}

\begin{document}
%%%%%%%%%%%%%%%%%%%%%%%%%%%%%%%%%%%%%%%%%%%%%%%%%%%%%%%%%%%%%%%%%%
%                        proctext.tex                            %
% Sample file for the proceedings of the conference in           %
% Imperial College, 1996. Adapted from sprocl.tex (from          %
% World Scientific Publishing Co) by Kris Thielemans.            %
%                                                                %
%%%%%%%%%%%%%%%%%%%%%%%%%%%%%%%%%%%%%%%%%%%%%%%%%%%%%%%%%%%%%%%%%%

% A useful Journal macro
%\def\Journal#1#2#3#4{{#1} {\bf #2}, #3 (#4)}
\newcommand{\Journal}[4]{{#1} {\bf #2}, #3 (#4)}
% Some useful journal names
\newcommand{\NCA}{\em Nuovo Cimento}
\newcommand{\NIM}{\em Nucl. Instrum. Methods}
\newcommand{\NIMA}{{\em Nucl. Instrum. Methods} A}
\newcommand{\NPB}{{\em Nucl. Phys.} B}
\newcommand{\PLB}{{\em Phys. Lett.}  B}
\newcommand{\PRL}{\em Phys. Rev. Lett.}
\newcommand{\PRD}{{\em Phys. Rev.} D}
\newcommand{\ZPC}{{\em Z. Phys.} C}
\newcommand{\MPLA}{{\em Mod. Phys. Lett.} A} 
\newcommand{\APH}{\em Ann. Phys.} 

\newcommand{\ra}{\rightarrow}
\newcommand{\ko}{K^0}
\newcommand{\be}{\begin{equation}}
\newcommand{\ee}{\end{equation}}
\newcommand{\bea}{\begin{eqnarray}}
\newcommand{\eea}{\end{eqnarray}}
\newcommand{\R}{{\rm R}}
\newcommand{\F}{{\rm F}}

%Replace the title, authors, and addresses within the curly brackets
%with your own title, authors, and addresses; please use
%capital letters for the title and the authors. 

\title{WORLD--MANIFOLD AND TARGET SPACE ANOMALIES IN HETEROTIC
GREEN--SCHWARZ STRINGS AND FIVE--BRANES}

\author{ K. LECHNER and M. TONIN }
\maketitle
\vskip0.5truecm
\centerline{Dipartimento di Fisica, Universit\`a di Padova} 
\centerline{Istituto Nazionale di Fisica Nucleare, Padova, Italy}
\vskip0.5truecm

\centerline{Seminar given at the Conference {\it Gauge Theories, Applied
Supersymmetry}}
\centerline{{\it and Quantum Gravity}, Imperial College, London, July 1996}
\vskip0.5truecm
\centerline{{\it presented by} K. LECHNER}

%%%%%%%%%%%%%%%%%%%%%%%%%%%%%%%%%%%%%%%%%%%%%%%%%%%%%%%%%%%%%%%%
% You may repeat \author \address as often as necessary.       %
% DO NOT use \footnote in \talk or \author. If necessary, use  %
% \thanks instead.
% The line with \presentedby is optional. It will print a line %
%    Presented by name                                         %
%%%%%%%%%%%%%%%%%%%%%%%%%%%%%%%%%%%%%%%%%%%%%%%%%%%%%%%%%%%%%%%%
\begin{abstract}
The quantum consistency of sigma--models describing the
dynamics of extended objects in a curved background
requires the cancellation of their
world--volume anomalies, which are conformal anomalies for the heterotic 
string and $SO(1,5)$ Lorentz--anomalies for the heterotic five--brane, and
of their ten dimensional target space anomalies.
In determining these anomalies in a $D=10$ Lorentz--covariant back--ground
gauge we find that for the heterotic string the worldvolume anomalies 
cancel for 32 heterotic fermions while for the conjectured heterotic
five--brane they cancel for only 16 heterotic fermions, this result being
in contrast with the string/five--brane duality conjecture. For what 
concerns the target space anomalies 
we find that the five--brane eight--form Lorentz--anomaly polynomial 
differs by a
factor of $1/2$ from what is expected on the basis of duality. Possible
implications of these results are discussed.
\end{abstract}

\section{Introduction}

It is by now clear that
$p$--branes are fated to play a central role in the duality relations 
occurring in string theories and $M$--theory. Among these dualities an
interesting one is the heterotic string/heterotic five--brane
strong--weak--coupling duality \cite{DUAL}. 
Unfortunately until now no consistent five--brane theory, based on a 
classical action triggering also its quantum dynamics, does exist.
The principal problems are the following.

\noindent
1) Whereas there exists a $\kappa$--invariant classical action for the 
gravitational sector of the five--brane, for its heterotic sector
no $\kappa$--invariant classical action is known.

\noindent
2) The heterotic five--brane sigma model appears to be power counting 
non--renormalizable.

\noindent 
3) Is there a ten--dimensional space--time interpretation for the 
physical modes of the gravitational sector of the heterotic five--brane?
These physical modes are four fermionic plus four bosonic modes which
do not span a representation of $SO(8)$, the little ten--dimensional 
Lorentz group.

\noindent
4) What is the quantum heterotic five--brane? A classification of the 
six--dimensional topologies is not available, and, moreover 
a term like $\int{\sqrt{g}}
R\varphi$, which furnishes in the case of the NSR--string
the quantum expansion parameter, seems not available in the case of 
Green--Schwarz (GS)--extended objects.

\noindent
5) How many fermions are there in the heterotic sector?

\noindent 
6) Do the anomalies in the heterotic five--brane cancel?

\noindent
7) Can the resulting heterotic five--brane be dual to the heterotic
string?

The problems 1) -- 4) will not be addressed in this talk. 
For what concerns 1), if one chooses
fermions as basic fields for the heterotic sector and constructs a simply
minded action -- for example introducing a minimal coupling with the external
gauge fields -- $\kappa$--invariance is destroyed. 

For what concerns renormalizability, point 2), from a dimensional point of
view the theory, living in six dimensions, does not seem renormalizable;
but if eventually a $\kappa$--invariant formulation will be found 
it is possible that
$\kappa$--invariance prevents the appearance of non--renormalizable
divergences in the effective action.
A similar conjecture has, in fact, been made for the eleven dimensional
membrane by Paccanoni {\it et al} \cite{PPT}.
Actually, the analysis of this paper is not complete
since an exhaustive classification of all the possible divergences
is very difficult to realize and has not been made. On the other hand,
GS--strings conformal invariance, which is fundamental
for its quantum consistency, is entailed by $\kappa$--invariance 
and it may be that for five--branes, and other GS--extended objects,
$\kappa$--invariance is just as fundamental for their quantum consistency
as conformal invariance is for strings.

The points 5) -- 7) will be addressed in this talk  and we concentrate on the
possible quantum consistency of the heterotic super--five--brane sigma model
embedded in an $N=1$, $D=10$ target superspace, i.e. on the derivation and
cancellation of its anomalies. This analysis will give us a concrete information
on the field content of the heterotic sector and shed some new unexpected
light on string/five--brane duality. The results presented in this talk
have been obtained in refs. \cite{LT1,LT2} to which we refer the reader
for the details of their derivation and for more detailed references.

Since $p$--brane sigma models are defined by GS--type actions,
like the GS--heterotic string, a natural attempt 
in the five--brane anomaly analysis consists in trying to extend, 
as much as possible,
the techniques we use in GS--string theory 
to the five--brane sigma model. We will start with the world--volume
anomalies which are conformal anomalies for the string and $SO(1,5)$
local Lorentz anomalies for the five--brane. Actually, for the GS heterotic
string
the conformal anomaly is cohomologically tied to the $SO(1,1)$ local
Lorentz anomaly and the $\kappa$--anomaly, while for the five--brane the
$SO(1,5)$ anomaly is cohomologically tied to the $\kappa$--anomaly, via the
Wess--Zumino consistency conditions. This means that it is sufficient to
worry about $SO(1,1)$ ($SO(1,5)$) anomalies only: once they cancel
all other worldsheet (worldvolume) anomalies will cancel automatically.
So first of all we have to have a good understanding of the $SO(1,1)$ 
anomaly cancellation in the string. 
This leads us to face the "conformal anomaly puzzle" i.e. a naif counting of 
the chiral fermionic degrees of freedom in the GS heterotic string 
leaves a non vanishing anomaly there: the left handed $\vartheta$--fields 
are 16 which by $\kappa$--symmetry are reduced to 8, while the right
handed heterotic fermions are 32, so the $\vartheta$'s are by a factor of
4 to short to cancel the $SO(1,1)$ anomaly. For what concerns the conformal 
anomaly in the non--supersymmetric sector,
the $X^m$ plus $(b,c)$--fields count $10-26=-16$ while the 
$\vartheta$'s count ${1\over 2}\cdot 8=4$ and again their contribution should
in some way be multiplied by 4 to lead to a cancellation. 
The conformal anomaly cancellation mechanism for the GS--string 
has been discovered in the flat case, in a $D=10$ Lorentz
covariant background gauge, by Wiegmann \cite{Wieg}. Our procedure for 
the $SO(1,1)$ anomaly cancellation 
mechanism in the sigma--model is based on this paper. 

The cancellation of the target space anomalies (via the GS--mechanism)
is necessary for the quantum consistency of the string/five--brane
sigma--models since they are cohomologically tied to genuine sigma--model
worldvolume $\kappa$--anomalies \cite{LT1,LT2,CLT} 
i.e. which vanish only in the flat limit.\footnote{A part from these
one expects additional genuine sigma--model $\kappa$--anomalies which are
$SO(1,9)$ and $SO(1,1)/SO(1,5)$ invariant and can be cancelled by modifying
the target superspace constraints \cite{CLT,T}, which we do not consider
here.}

In section 2 we will show that the above mentioned  quadruplication is 
intimately related
to the target space $SO(1,9)$ local Lorentz anomaly in the GS--string,
and obtain the expected complete
four--form anomaly polynomial for the heterotic string. Encouraged by this
result we will in section 3 extend this 
method to the five--brane sigma model and compute its complete 
eight--form anomaly polynomial, under the assumption that the heterotic
sector is made out of a certain number of fermions. Our principal results
are the following. The number of fermions needed to cancel the worldvolume
anomaly is sixteen rather than the expected thirtytwo.
On the other hand 
the coefficient of the $D=10$ target space Lorentz anomaly carries a 
factor of $1/2$ with respect to what is expected on the basis of duality.
Section 4 is devoted to a brief discussion of our results.

\section{Heterotic string anomalies}

The sigma--model action for the heterotic Green--Schwarz string 
with gauge group $SO(32)$ in ten target space--time dimensions 
is given by
\be
S_2 = - {1 \over 2\pi \alpha'} \int d^2 \sigma \left({1\over2}
\sqrt{g}\ g^{i j} V_i^a V_{j a}
+\widetilde{B_2}
- {1\over2}\sqrt{g}\ e_-^j \psi 
(\partial_j - A_j) \psi \right). 
\label{az2}
\ee
Here the string fields are the supercoordinates $Z^M = (X^m, 
\vartheta^\mu )$, the 32 heterotic fermions $\psi $ and the
worldsheet zweibeins $e_{\pm}^i, g^{ij} = e_+^{(i} e_-^{j)}$.
The induced zehnbeins are 
given by $V_i^A = \partial_i Z^M E_M{}^A (Z)$ and $\widetilde{B_2}$ is 
the pullback on the string worldsheet of the supergravity two--superform 
$B_2$. 

This action is invariant under $d=2$ diffeomorphisms, local 
$SO(1,1)$ Lorentz transformations, conformal and $\kappa$--transformations.
Diffeomorphisms anomalies can always be eliminated at the expense of 
conformal/$SO(1,1)$ anomalies, so we will not dwell upon them. Since the
coefficient of the conformal and $\kappa$--anomalies is tied for
cohomological reasons to the $SO(1,1)$ anomaly we will now concentrate
on the last one. Since this is an ABBJ--anomaly only fermions will
contribute, in our case the $\vartheta$'s and the $\psi$. The contribution
of the latters is standard, so we will now consider in detail the
formers. It is most convenient to use the background field method
together with a normal  coordinate expansion; calling the quantum 
$\vartheta$'s $y^{\alpha}$ where $\alpha= 1,\cdots,16$ the relevant part
of the expanded action becomes
\be
I (V, \Omega, y) = {1 \over 2} \int d^2 \sigma \sqrt{g}\ g^{ij} V^a_i \
y\ \Gamma_a 
{1 - \Gamma \over 2} D_j {1 - \Gamma \over 2} \ y 
\label{ae}
\ee
where $D_j \equiv \partial_j - {1 \over 4} \Gamma{}_{cd} 
\Omega_j{}^{cd}$,  $\Omega_j{}^{cd}$ is the $SO(1,9)$ target space 
Lorentz connection, the $\Gamma^a$ are ten dimensional Dirac matrices
and we defined the matrix
$
\Gamma^\alpha{}_\beta = {1\over V_+^a V_{a-}} \cdot{\varepsilon^{i j} \over 
\sqrt{g}}  V^a_i V^b_j (\Gamma_{ab})^\alpha{}_\beta.
$
An  $SO(1,9)$--covariant background gauge fixing can now be achieved
by imposing ${1 + \Gamma \over 2} \ y= 0$,
which reduces the physical $y$'s from 16 to 8,
but the problematic feature of (\ref{ae}) is that the kinetic term for the 
$y$'s is not canonical in that it is multiplied by the external (classical)
fields $V_i^a$ and one can not define a propagator. Eq. (\ref{ae}) can be
transformed to an action with a canonical kinetic term, taking advantage
from its manifest classical $SO(1,9)$ invariance, by applying
a convenient $SO(1,9)$ Lorentz rotation with group element 
$\Lambda_{a}{}^b$. But, since
the integration measure $\int \{{\cal D} y\}$ under local 
$SO(1,9)$ transformations is not invariant \cite{CLT}, this rotation
gives in general rise to a Wess--Zumino term. The $SO(1,9)$ Lorentz 
anomaly, contrary to the $SO(1,1)$ anomaly, can be computed with  standard
techniques and the corresponding polynomial turns out to be \cite{CLT}
\be
X_L^{(2)}={1\over 8\pi}tr\R^2\equiv {1\over 8\pi} \ d\omega_3(\Omega),
\label{L2}
\ee
where $R_a{}^b$ is the $D=10$ Lorentz curvature two--form and 
$\omega_3(\cdot)$ is the standard Chern-Simons three--form. Therefore, 
for a generic rotation, $\Lambda$, the measure $\int \{{\cal D} y\}$, and
hence the effective action, change by a Wess--Zumino term given by
\be
\Gamma_{WZ}={1\over 8\pi}\int_{D_3}\left(\omega_3(\Omega)-
\omega_3(\Omega^\Lambda)\right),
\label{WZ}
\ee
where the boundary of $D_3$ is the worldsheet.
The crucial point is that for
the particular $\Lambda_a{}^b$ which renders the kinetic term of
the $y$'s canonical \cite{LT1} one has 
$\omega_3(\Omega^\Lambda)=\omega_3(\omega^{(2)}) + Y_3+dY_2,$
where $\omega^{(2)}$ is the {\it two}--dimensional Lorentz connection,
$Y_2$ is a local form and can therefore be disregarded and $Y_3$ is 
an $SO(1,1)$ {\it and} $SO(1,9)$--invariant form. The Wess--Zumino term
(\ref{WZ}) contributes therefore to the $SO(1,1)$ anomaly with a
polynomial which is given by
\be
X_{WZ}^{(2)}=-{1\over 8\pi}tr{\cal R}^2=-{1\over 192\pi}\cdot 
24 \ tr{\cal R}^2,
\label{WZA2}
\ee
where ${\cal R}$ is the two--dimensional Lorentz curvature 
two--form (all traces are in the fundamental representations of the 
orthogonal groups). The 
functional integral over the (transformed) $y$'s is now canonical and
corresponds to eight  Weyl--Majorana fermions with
effective action given by 8 $\ell n \det ^{1/2}(\sqrt{g}\ \partial_+)$;
this entails a contribution to the anomaly given by \cite{AGG}
\be
X^{(2)}_{naif}=-{1\over 192\pi}\cdot 8 \ tr{\cal R}^2.
\label{A02}
\ee
The total contribution of the quantum $\vartheta$'s to $SO(1,1)$ and 
$SO(1,9)$ anomalies is thus obtained by 
summing up (\ref{L2}),(\ref{WZA2}) and (\ref{A02}):
\be
X^{(2)}_\vartheta={1\over 2\pi}\left(-{8+24\over 96}tr{\cal R}^2
+{1\over 4}tr\R^2\right).
\label{vartheta2}
\ee
We see that the Wess--Zumino term leads to a quadruplication of the 
"naif" $SO(1,1)$  anomaly. 

The contribution of $N_\psi$ right--handed heterotic Majorana--Weyl 
fermions, which contribute only to $SO(1,1)$ and Yang--Mills anomalies,
can be read directly from the index theorem \cite{AGG}, 
$
X^{(2)}_\psi={1\over 2\pi}\left({N_\psi\over 96}tr {\cal R}^2
-{1\over 4}tr \F^2\right).
$
Summing up this and (\ref{vartheta2}) we obtain the total worldsheet and
target space anomaly polynomial for the heterotic string as
\be
X^{(2)}={1\over 2\pi}\left({N_\psi-(8+24)\over 96} \ tr{\cal R}^2
+{1\over 4}\left(tr\R^2-tr\F^2\right)\right).
\label{A2}
\ee
The worldsheet anomaly cancels for 32 heterotic fermions, the gauge group
can therefore be taken to be $SO(32)$ and the remaining target space anomaly
can be cancelled by modifying the $B_2$ Bianchi identity to
\be
dH_3=-2\pi\alpha'\cdot{1\over 8 \pi}\left(tr\R^2-tr\F^2\right)
\equiv -2\pi\alpha'\cdot I_4,
\label{I4}
\ee
in agreement with the GS mechanism.

\section{Heterotic five--brane anomalies}

The action for the 
super--fivebrane sigma--model \cite{BTS} embedded in an $N=1$, $D=10$
target space supergravity background  is given by 
\be
S_6=-{1\over(2\pi)^3\beta^\prime} \int d^6\sigma
\left( {1\over 2} e^{-{2\over 3}\varphi}\sqrt{g} g^{ij} V_i^a V_{ja} 
-\widetilde {B_6} -2 \sqrt{g}\right),
\label{az6}
\ee
where $\widetilde{B_6}$ is the pullback on the six--dimensional
worldvolume of the dual supergravity six--superform $B_6$. $S_6$
is invariant under $\kappa$--transformations, $d=6$ diffeomorphisms
and $SO(1,5)$ local Lorentz transformations if one replaces the
metric $g_{ij}$ with sechsbeins. As in the case of the string it is
sufficient to worry about $SO(1,5)$ and $SO(1,9)$ anomalies only. 
As we will see, the action in eq. (\ref{az6}) will give rise to a 
non--vanishing $SO(1,5)$ anomaly, therefore one  {\it must} add a
heterotic sector to cancel this anomaly.
Despite the difficulties mentioned in the introduction
we will assume that this sector is made out of a certain number $N_\psi$
of $d=6$ complex Weyl fermions, minimally coupled to Yang--Mills fields
of a gauge group $G$. A part from this, the derivation of the anomalies 
follows mainly the strategy we adopted in section 2 for the string, 
so we will only report the results referring to \cite{LT2} for the
details of their derivation.

The total $SO(1,5)$ and $SO(1,9)$ anomaly 
due to the $\vartheta$'s is 
again a sum of three terms, like (\ref{L2}),(\ref{WZA2}) and (\ref{A02}),
$X^{(6)}_\vartheta= X_L^{(6)}+X_{WZ}^{(6)}+X_{naif}^{(6)}$,
and the formula analogous to (\ref{vartheta2}) is 
\bea
X^{(6)}_\vartheta &=&
{1\over 192(2\pi)^3}  
  \left( ( -1-15)\left({1\over 30}tr {\cal R}^4 + {1\over 24}
\left(tr {\cal R}^2\right)^2\right)\right.\nonumber\\
& &\left.+ \ tr {\cal R}^2 tr \R^2 - {3\over 8} 
\left(tr \R^2\right)^2 + 
{1\over 2}tr \R^4\right).
\label{vartheta6}
\eea
In this case the Wess--Zumino term (counting for 15 complex Weyl fermions)
amounts to multiply the naif $SO(1,5)$ 
anomaly (corresponding to 1 fermion, i.e. the 8 physical real $\vartheta$'s)
by a factor of 16. The index theorem gives for the heterotic fermions,
with chirality opposite to that of the $\vartheta$'s,
\be
X_\psi^{(6)}=
{1\over 192(2\pi)^3}
\left( N_\psi\left({1\over 30}tr {\cal R}^4 + {1\over 24}
\left(tr {\cal R}^2\right)^2\right)
-2\ tr \F^2  tr{\cal R}^2 +8\ tr \F^4
\right).
\label{Het6}
\ee
The total heterotic five--brane anomaly, which is gotten summing up 
(\ref{vartheta6}) and (\ref{Het6}), becomes:
\bea
X^{(6)}&=&
 {1\over 192 (2\pi)^3} \left((N_\psi-1-15) \left( {1\over 30}tr
{\cal R}^4 +{1\over 24} \left(tr {\cal R}^2\right)^2\right)\right.
\nonumber\\
& &+ \left(2 tr {\cal R}^2 -
tr \R^2\right) \left({1\over 2}tr \R^2 -  tr\F^2\right)
\nonumber\\
& & \left.  + {1\over 2}\left(tr \R^4 + {1\over 4} 
\left(tr \R^2\right)^2\right) -  tr \F^2 tr \R^2 + 8 tr
\F^4\right).
\label{A6} 
\eea

\section{Discussion}

One aspect of the string/five-brane duality conjecture emerges from the 
factorization of the $N=1$, $D=10$ supergravity anomaly polynomial,
$I_{12}={1\over 2\pi}I_4\cdot I_8$, where for the gauge group $SO(32)$
$I_4$ is given in eq. (\ref{I4}) and 
\be
I_8 
= {1\over 192 (2\pi)^3} \left(tr\R^4 + {1\over 4}
(tr \R^2)^2 - tr \R^2 tr{\cal F}^2 + 8\ tr {\cal F}^4\right),
\label{I8}
\ee
where the Yang--Mills curvature ${\cal F}$ belongs to the fundamental
representation of $SO(32)$. According to the conjecture, once in (\ref{A6})
the worldvolume anomaly cancels, the remaining
target space anomaly polynomial should coincide with (\ref{I8}). To 
cancel the worldvolume anomaly
one needs $N_\psi=16$, i.e. sixteen heterotic fermions, and therefore the 
gauge group can not be $SO(32)$ (and not even $E_8\otimes E_8$) and one can 
not identify $\F$ with ${\cal F}$. Moreover, there are mixed terms in
(\ref{A6}), $2 tr {\cal R}^2 \cdot\left({1\over 2}tr \R^2 -  tr\F^2
\right)$, which can be cancelled in no way, and the weights of the
leading target space Lorentz anomaly, $tr\R^4$, in $X^{(6)}$ and
$I_8$ differ by a factor of $1/2$. 

To quantify these discrepancies let us assume that the $\vartheta$'s count 
for {\it two}, instead of one, complex Weyl fermions. In this case the
total anomaly polynomial would be given by 
$\widetilde{X^{(6)}}=2\cdot X^{(6)}_\vartheta+X^{(6)}_\psi$ which can be written
as 
\be
\widetilde{X^{(6)}}=I_8+{1\over 48(2\pi)^2}\left(2 tr {\cal R}^2 -
tr \R^2\right)\cdot I_4
+{N_\psi-32\over 192(2\pi)^3}\cdot
\left({1\over 30}tr {\cal R}^4 + {1\over 24}
\left(tr {\cal R}^2\right)^2\right).
\ee
In this case one would need 32 heterotic fermions in the fundamental 
representation of $SO(32)$, the term proportional to $I_4$ would correspond 
to a trivial anomaly thanks to (\ref{I4}), and $\widetilde{X^{(6)}}$ would
reduce to $I_8$ -- in complete agreement with duality -- which could be 
eliminated by modifying the Bianchi identity of $B_6$ to
$dH_7=(2\pi)^3\beta^\prime I_8$. Since, according to duality, $H_7$ has
to be the Hodge--dual of $H_3$ this Bianchi identity, together with 
(\ref{I4}), would imply a relation between the charges of strings
and five--branes involving the ten--dimensional Newton's constant
$\kappa$, i.e. $2\kappa^2=(2\pi)^5\alpha^\prime\beta^\prime$, which 
corresponds to a Dirac--like quantization condition \cite{NT} with $n=1$.

So our principal conclusion is that the five--brane $\vartheta$--anomaly 
is only half 
of what is expected on the basis of string/five--brane duality, adding a new
problem to the ones already mentioned in the introduction. We can 
nevertheless mention that if we set in $X^{(6)}$ and $I_8$ the gauge fields
to zero, ${\cal F}=\F=0$, then the worldvolume anomaly cancels for
sixteen heterotic fermions and by subtracting a suitable trivial anomaly,
as above, $X^{(6)}$ would reduce to  ${1\over 2}\cdot I_8$. This would imply
the quantization condition $2\kappa^2={1\over 2}\cdot
(2\pi)^5\alpha^\prime\beta^\prime$ i.e. $n={1\over 2}$ which signals
the presence of half--charged five--branes. 
Half--charged fivebranes arose, actually, in ref. \cite{Pol}
where they appear, however, always in pairs such that their total
charge is always integer. Half integral magnetic charges have arisen
also on fixed points of $Z_2$-orbifold compactifications of $N=1,D=11$
Supergravity in ref. \cite{HW}.

%\section*{References}

\end{document}